\documentclass[12pt, preprint]{aastex}
\usepackage{epsfig}

\newcommand{\saxj}{\mbox{SAX J1808.4$-$3658}}

\begin{document}
\title{What Makes an Accretion-Powered Millisecond Pulsar?} 
\author{Feryal \"{O}zel}
\affil{Departments of Physics and Astronomy, University of Arizona,
1118 E. 4th St., Tucson, AZ 85721; fozel@physics.arizona.edu}

\begin{abstract}

We investigate the dependence of pulse amplitudes of accreting
millisecond pulsars on the masses of the neutron stars. Because the
pulsation amplitudes are suppressed as the neutron stars become more
massive, the probability of detection of pulsations decreases in
systems that have been accreting for a long time. However, the
probability of detectable pulsations is higher in transient systems
where the mass accretion is sporadic and the neutron star is likely to
have a low mass. We propose this mechanism as the explanation of the
small number of millisecond X-ray pulsars found to date, as well as
their emergence as fast pulsars mostly in transient, low-$\dot M$
systems. This mechanism can also quantitatively explain the lack of
pulsars in the majority of LMXBs.

\end{abstract}


\section{Introduction}

The scarcity of accreting neutron stars with detectable millisecond
spin frequencies has been a puzzle since the suggestion of Alpar et
al.\ (1982) and Radhakrishnan \& Srinivasan (1982) that the
millisecond radio pulsars are spun-up remnants of neutron stars in
low-mass X-ray binaries (LMXBs). Given the magnetic field distribution
of recycled radio pulsars between $10^{8-9}$~G, these progenitors are
expected to possess a field of comparable strength, which plays a
dynamically important role in the accretion process. In particular,
accretion is likely to proceed by channeling matter onto the polar
caps, which can then result in modulations in the X-ray lightcurves of
the accreting neutron stars that can be detected as millisecond X-ray
pulsars (hereafter ms~pulsars). However, deep searches until 1998
yielded no sign of these progenitors.

The discovery of a small number of ms pulsar sources (e.g., Wijnands
\& van der Klis 1998; Markwardt et al.\ 2002; Galloway et al. 2002,
2005) gave the first solid support to the idea of recycling. To date,
there are eight accreting neutron stars that show periodic modulations
in their persistent X-ray lightcurves, with frequencies ranging
between 182 and 599~Hz (see Table 2; see also the discussion for
detections of intermittent pulsations in the bulk LMXB
population). These discoveries, at the same time, raised numerous
questions, two of which are important for the present paper: What
conditions lead to the generation of a ms pulsar?  How do these
sources differ from the remaining population of LMXBs?

A potential clue into their nature comes from the fact that companions
of all the discovered ms pulsars have peculiar properties that
distinguish them from the general population of LMXBs, of which $\sim
150$ are known (Liu et al. 2007). Specifically, the companions have
extremely low masses and are found in tight binary orbits
(40~min$-$4~h), where mass transfer is gravitationally driven,
resulting in very low time-averaged mass accretion rates. This is
corroborated by the X-ray observations of their weak and sporadic
outbursts. If the mass accretion rates have remained low over the
lifetime of the binary, then the overall effect is a very small amount
of mass accreted onto the neutron star.

Several possible explanations have been proposed for the absence of
pulsations from the bulk of the LMXB population (Brainerd \& Lamb
1987; Kylafis \& Phinney 1988; Mezsaros, Riffert, \& Berthiaume 1988;
also see G\"o\u{g}\"u\c{s}, Alpar, \& Gilfanov 2007). None of the
mechanisms rely on, or explain, the fact that all ms pulsars have been
observed only in particular types of binary systems. The growing
number of ms pulsars known today make it highly unlikely that this is
coincidental. The low inferred time-averaged mass accretion rates led
to one possible explanation that the magnetic fields in ms pulsars do
not decay to as low values as they do in the general LMXB population
(Cumming, Zweibel, \& Bildsten 2001) and hence can channel the
accretion flow and produce visible pulsations. Within this scenario,
it is difficult to understand the striking similarities in the spectra
and broadband timing properties of the ms pulsars and the LMXBs when
they have similar accretion rates (Psaltis \& Chakrabarty 1998), as
well as the detection of intermittent pulsations from two LMXBs.

In this paper, we propose an alternative explanation for the presence
of detectable pulsations in this small number of binaries. As neutron
stars accrete matter, their masses increase while, for most of the
neutron-star matter equations of state, their radii remain
approximately constant (or even decrease). This results in a rapid
increase in the compactness $GM/Rc^2$ of the neutron stars and very
efficiently suppresses the amplitudes of pulsations due to
gravitational light bending. When all binary and observer
configurations are taken into account, the statistical probability of
detecting pulsations from such a source thus decreases rapidly as the
amount of accreted mass on the neutron star increases. We show that
this mechanism can explain quantitatively the lack of pulsations in
the majority of LMXBs that have been accreting at high rates while
allowing for ms pulsations to be detected from the low accretion rate
systems.

\section{Pulse Amplitudes of Accreting Neutron Stars}

Pulsations originating from the surface of a neutron star are subject
to strong gravitational lensing. General relativistic bending of light
in the strong stellar gravitational field of the neutron star
suppresses the modulations of the observed lightcurves (Pechenick,
Ftaclas, \& Cohen 1983). For rapidly rotating neutron stars, time
delays and Doppler boosts also affect the lightcurves seen by an
observer at infinity. In such cases, both of these phenomena render
the lightcurves asymmetric and more peaked, giving rise to higher
amplitudes of pulsations both at the spin frequency and its harmonics
(Braje, Romani, \& Rauch 2000; Weinberg, Miller, \& Lamb 2001; Muno,
\"Ozel, \& Chakrabarty 2002).

The flux modulations observed in the lightcurves of accreting X-ray
pulsars are attributed to the hot polar caps where the accretion
columns hit the stellar surface. Therefore, all these relativistic
effects play a role in determining the pulsation amplitudes and thus
the observability of ms~pulsars. To calculate the trajectories of the
photons reaching the observer taking into account light bending, time
delays and Doppler boosts, we use the numerical code described in Muno
et al.\ (2002).

Our primary goal is to investigate the dependence of pulsations on the
properties of the neutron star, and in particular, on its increasing
mass as accretion proceeds. However, a number of other phenomena
affect positively or negatively the pulsation amplitudes of
ms~pulsars. For example, as the spin frequency $\omega$ increases, so
does the opening angle of the accretion column (see below), which
reduces the amplitudes. On the other hand, Doppler and time delay
effects, which increase the pulsation amplitudes, are more pronounced
in fast rotators. In our present examples, we take into account both
of these effects when we specify the properties of the neutron star
systems.

We assume that the accretion onto the millisecond pulsars proceeds via
a geometrically thin disk that is truncated near but inside the outer
corotation radius. At this ``loading'' radius, matter is attached onto
and follows the field lines onto the neutron star surface. A large
fraction of the X-ray emission, especially in the soft X-rays, is
thought to originate at localized polar caps on the stellar surface
where the accretion columns terminate. We assume that the polar caps
are circular and denote their opening angle by $\rho$. The corotation
radius is given by 
\begin{eqnarray} R_{\rm co} &\equiv&
(GMP^2/4\pi^2)^{1/3} \\
           &\simeq& 3.5R \left(\frac{R}{10\,{\rm km}}\right)^{-1} 
                         \left(\frac{M}{1.4\,M_\odot}\right)^{1/3}
                         \left(\frac{P}{3\,{\rm ms}}\right)^{2/3}.
\end{eqnarray}
We estimate the opening angles of each polar cap by noting that the
quantity $\sin^2 \theta/r$ remains constant along the field lines,
yielding
\begin{equation}
\frac{\sin^2 \rho}{R}=\frac{\sin^2 90^\circ}{R_{\rm co}},
\end{equation}
and thus $\rho \simeq 35^\circ$.  Therefore, we use two antipodal
regions of $\rho=40^\circ$ when calculating the amplitude of
pulsations from the millisecond X-ray pulsars. Clearly, for a variable
mass accretion rate on a given source, the corotation radius, the
loading radius, and, thus, the size of the polar caps are also subject
to variations.

Another important factor that affects the amplitudes of pulsations is
the angular dependence (beaming) of the emission from the poles and
the accretion column. The angular dependence is unique to the emission
process and can be affected by variables such as the magnetic field
strength or the temperature of the poles. The spectra of the
millisecond pulsars suggest that the $2-20$~keV emission results from
the Comptonization of the photons that are emitted at the boundary
between the accretion column and the stellar surface. In the present
calculations, we use a Hopf function to describe the beaming of
emission, which is appropriate for energy-independent scattering in
unmagnetized media. This can be justified for the following two
reasons. First, the fact that the timing properties are largely
independent of the photon energy in the observed energy range (e.g.,
Cui et al.\ 1999) indicate that in this range, all photons have
experienced multiple scatterings and that the energy change in the
scatterings probably has little effect on the angular
distribution. Second, Psaltis \& Chakrabarty (1999) estimate the
magnetic field strength of \saxj\ to be $B \sim 10^8-10^9$~G based on
timing arguments. This yields an electron cyclotron energy that is
much below the photon energy range of the observations, and therefore
the magnetic field has no effect on the properties of the emission. We
note that detailed modeling of the pulse profile and the source
variability also point to Comptonization in the accretion columns and
yield similar beaming functions as the one used here (Poutanen \&
Gierlinski 2003). 

\begin{figure}
\centerline{\epsfig{file=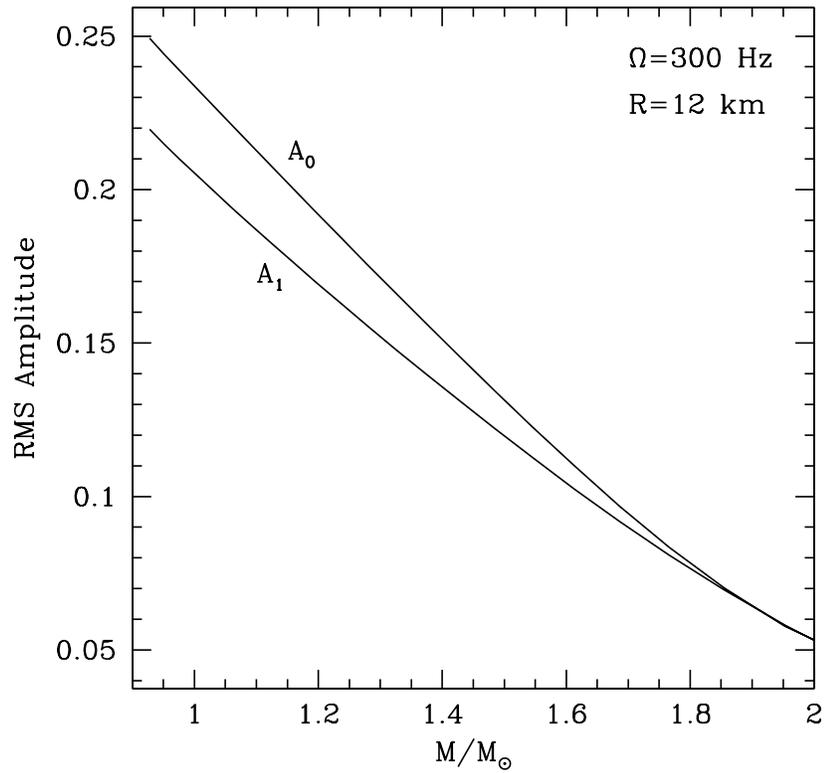,width=0.75\linewidth}}
\caption {The dependence of the pulse amplitudes on the mass of the
accreting neutron star with spin frequency $\Omega=300$~Hz. $A_0$ and
$A_1$ refer to the fundamental and first harmonic rms amplitudes,
respectively. Accretion columns are assumed to have an opening angle
of $\rho \approx 40^\circ$ on the stellar surface.  Here, the observer
as well as the magnetic poles are located at $\beta=\alpha=60^\circ$
from the rotation axis. }
\label{Fig:ampl}
\end{figure}

Figure~\ref{Fig:ampl} shows the dependence of the fundamental and
first harmonic rms pulse amplitudes on the mass of the accreting
neutron star, assuming a constant stellar radius of $R=12$~km and for
a rotational frequency of $\Omega=300$~Hz. Note that for most
equations of state of neutron star matter, there is a characteristic
mass-radius relationship such that the stellar radius remains constant
for a large range of mass values (e.g., Glendenning 2000). Therefore,
as the neutron star accretes matter, its compactness ratio $GM/(Rc^2)$
increases and leads to a rapid suppression of pulsation amplitudes
(e.g., Pechenick, Ftaclas, \& Cohen 1983; \"Ozel 2002). The spin
frequency of 300~Hz is a representative value for the range
$182-599$~Hz observed to date in ms~pulsars. This figure demonstrates
that the rms amplitudes are suppressed by a factor of $2-3$ for an
increase in the stellar mass of about $0.4 M_\odot$. Such a drastic
suppression is also seen for a wide range of values for the stellar
radii and spin frequencies, as well as the thicknesses of the
accretion columns.

The apperance of the pulsations depends strongly on the two angles
$\alpha$ and $\beta$ that specify the inclination of the magnetic axis
and the line-of-sight of a distant observer with respect to the
rotation axis, respectively. In Figure~\ref{Fig:ampl}, we chose an
average set of values of $\alpha = \beta = 60^{\circ}$ for these
angles. In the next section, we present a statistical treatment for
the entire range of the magnetic inclination and the observer's
inclination and determine the fraction of binary systems that would be
visible as pulsars.

\section{Detection Probabilities}

We now address pulsation amplitudes statistically for low-mass and
high-mass accreting neutron stars. We calculate rms pulse amplitudes
taking into account a population of neutron stars uniformly
distributed in their geometrical quantities; i.e., their magnetic
inclinations and observer's inclinations. Our aim is to determine
the detection probabilities for these systems.

We keep other intrinsic neutron star properties fixed at values
specified in the previous section. For the neutron star mass-radius
relation, we choose an equation of state commonly referred to as UU
(see Cook, Shapiro, \& Teukolsky 1994), which {\it (i)} corresponds to
a radius of approximately 12~km in the mass range of interest {\it
(ii)} has a maximum mass of $2.2 M_\odot$ and can thus support the
heavier neutron stars formed by accretion and for which there is
recent observational evidence (Mu\~noz-Darias, Casares, \&
Martínez-Pais 2005; \"Ozel 2006; Cornelisse et al.\ 2007), and {\it
(iii)} is not excluded to date by any other probes of neutron-star
matter. Note that there are several alternative equations of state of
neutron star matter that give quite similar mass-radius relations for
the range of masses under consideration (e.g., Lattimer \& Prakash
2007).

In Figure~2, we show the fraction of systems that appear as a pulsar
with an amplitude given on the x-axis for two values of the neutron
star mass. As in the particular example shown in Figure~1, when the
neutron star mass is low, the pulse amplitudes are higher and a
significantly larger fraction of systems are visible as pulsars, with
amplitudes as large as $\sim 10\%$ in a few cases. When the mass is
higher, however, rms amplitudes are much lower, and the fraction of
systems that are detectable as pulsars with a given amplitude drops
rapidly.

Based on the accretion histories observed in these systems, we propose
a connection between low-mass ($m=M/M_\odot=1.4$) systems and the
accreting millisecond X-ray pulsars from which $\gtrsim 5\%$ amplitude
pulsations can be detected in roughly $50\%$ of the time. On the other
hand, the probability of detecting pulsations from a high-mass neutron
star, such as those that can emerge in a persistent, high $\dot{M}$
LMXB, drops to about $10\%$ for a $2\%$ pulse amplitude. Note that no
such system appears with a pulse amplitude $\gtrsim 2.5\%$.

\section{Discussion and Conclusions}

In this paper, we have attributed the difference between pulsing and
non-pulsing neutron stars in LMXBs to their masses. There are several
mechanisms that contribute to the evolution of binaries and determine
the mass transfer rate (see, e.g., Verbunt 1993). In binaries with
orbital periods greater than $\sim 1$~d, evolved giant companions lose
mass rapidly, giving rise to high mass accretion rates. On the other
hand, binaries with smaller orbital periods evolve because of loss of
orbital angular momentum. For companion stars more massive than
$\gtrsim 0.3~M_\odot$, the dominant magnetic braking mechanism can
drive mass losses of order $0.1 \dot{M}_{\rm Edd}$. For less massive
companions with $M \lesssim 0.3 M_\odot$, it is thought that the lack
of a convective envelope results in the evolution of the binary solely
due to the emission of gravitational waves. In this regime, time
averaged mass accretion rates are only $<\dot{M}> \approx 10^{-11}
M_\odot$~yr$^{-1}$, or less than $\sim 0.1\% \dot{M}_{\rm Edd}$.
Binaries harboring the ms pulsars most probably belong to this last
category, where very short binary periods point to very small mass,
degenerate, or even brown dwarf, companions (e.g., Bildsten \&
Chakrabarty 2001).

We looked for observational evidence of these different accretion
rates in the bulk LMXB population and the ms~pulsars using the {\it
Rossi} X-ray Timing Explorer (RXTE) All Sky Monitor
lightcurves. Members of both of these populations show outbursts on
the timescale of months, while a large number of LMXBs are persistent
sources. We integrated their X-ray countrates over the entire period
spanning January 1996 to December 2007 over which they have been
observed and converted these to X-ray fluxes taking into account the
fact that 1~Crab$\simeq 2 \times
10^{-8}$~erg~cm$^{-2}$~s$^{-1}=75$~ASM count~s$^{-1}$. Using distances
estimated from photospheric radius-expansion bursts for each source
(Galloway et al.\ 2008), we obtained a distribution of X-ray
luminosities and converted these values into time-averaged mass
accretion rates $<\dot{M}>$ (see Table~1). Figure~3 shows the
histogram of mass accretion rates for both populations. As expected
from theoretical arguments, the sources that appear as ms~pulsars
occupy the lowest end of the distribution of time-averaged accretion
rates, which span four orders of magnitude. This implies that
ms~pulsars on average indeed accrete much less mass than the average
LMXB, assuming that the trends observed by RXTE can be extrapolated
over the lifetimes of the binaries.

The histogram in Figure~3 includes only four of the eight ms~pulsars
and a subset of the LMXBs because the calculation of the accretion
rate requires a distance estimate to each source. Since we used
distance estimates based on photospheric radius expansion bursts,
which occur predominantly at low accretion rates, we expect that our
distribution underestimates the number of high $<\dot{M}>$-sources.
Correcting for this, however, would only accentuate the difference
between the pulsing and non-pulsing sources.

Differences in the mass accretion rates in the two populations may be
expected to lead to differences in their spin frequency distributions
if the latter are determined by magnetic spin equilibrium. If the
distributions in the magnetic field strength are similar between the
two populations, then the millisecond pulsars should have a spin
frequency distribution that is displaced towards lower values compared
to that of the non-pulsing LMXBs. Given the $\sim 2$~orders of
magnitude difference in the average mass accretion rates between the
two populations (Fig.~3), and a disk-magnetosphere interaction model
that best describes the pulse behavior of SAX J1808.4$-$3658 (the 1R
model in Psaltis \& Chakrabarty 1999 that has a $\dot{M}^{0.2}$
dependence), the mean spin frequency of the pulsing sources should be
a factor of $\sim 2.5$ lower than those of non-pulsing LMXBs. Figure~4
shows the spin frequency distributions of both populations together
with their median values. Indeed, the distribution of the pulsing
sources, with a median of 314~Hz, is systematically lower than that of
the non-pulsing sources that has a median of 550~Hz. 

An independent argument for the presence of massive neutron stars in
LMXBs comes from dynamical and spectroscopic mass measurements.
X1822$-$371 (Mu\~noz-Darias et al.\ 2005); Aql X-1 (Cornelisse et al.\
2007), 2S 0921-630 (Shahbaz et al.\ 2004; Jonker et al.\ 2005), and
EXO~0748$-$676 (\"Ozel 2006) all have inferred neutron star masses
exceeding $1.7~M_\odot$. Conversely, a detailed pulse profile modeling
of the ms~pulsar \saxj\ yields a neutron star mass of
$1.2-1.6~M_\odot$ (Poutanen \& Gierlinski 2003).

If the eight observed ms~pulsars are a representative sample of the
distribution over the observer's inclination and the magnetic
inclination, then the observed cumulative distribution of their pulse
amplitudes should follow the trend shown in Figure~2 down to a
limiting amplitude. Table~2 shows the rms amplitudes of the ms~pulsars
observed to date. In Figure~2, we compare the observed cumulative
distribution of pulse amplitudes, assuming that it is complete down to
1\%, to the theoretical expectation obtained for the set of
representative parameters discussed earlier. The remarkable
qualitative agreement provides additional support to the mechanism
proposed here.

This mechanism suggests that a fraction of the LMXB population should
also show millisecond pulsations at the level of $\sim 1\%$, assuming
no other processes play a role in determining pulsation amplitudes.
Current upper limits on the amplitudes of periodic oscillations in
LMXBs range from $\lesssim 1\%$ for the most luminous Z sources to a
few percent for the less luminous atoll sources. Most of these limits
come from studies with Ginga, which has been used to carry out the
only systematic search for the pulsations in LMXBs (see Vaughan et
al. 1994 and references therein). Eleven out of the fifteen sources
studied have very high mass accretion rates, which is likely to result
in very small rms amplitudes in our model. This is both because the
disk penetrates very close to the star at high mass accretion rates,
leading to very large cap opening angles, and also because the neutron
star can become very massive over its lifetime. Three out of the
remaining four sources have spin frequencies higher than the Nyquist
frequency of these observations (the timing resolution of this
instrument sets an upper limit to the detectable spin frequency of
$\sim 500$~Hz).


There have not been any published systematic studies with the Rossi
X-ray Timing Explorer to detect or place upper limits on the
pulsations of the LMXB sample statistically, which would serve as an
excellent test for the predictions we present here. However, there
have been isolated deep investigations, which yielded varying
results. One well-studied case is 4U~1820-30 (Dib et al.\ 2005) with
an upper limit of $0.8\%$ on the pulsation amplitude. The studies of
the numerous archival observations of two other LMXBs, in constrast,
resulted in the detections of pulsations: Aql X-1 showed pulsations in
a single, 150~s interval of data, with an amplitude of $1-10\%$
(Casella et al.\ 2008), while SAX J1748.9$-$2021 showed 442~Hz
intermittent pulsations with an amplitude of $2.1\%$ (Altamirano et
al.\ 2008; also see below for a discussion of intermittent
pulsations). This has led to the categorization of the latter source
as the ninth accretion powered ms~pulsar, with the largest orbital
period (8.9 hr) in that sample. However, given all its other
characteristics, within our model, SAX~J1748.9$-$2021 belongs to the
bulk of the LMXB population with higher mass accretion rates that
actually survives as a low amplitude pulsar.

While the general relativistic suppression of pulsation amplitudes for
neutron stars with higher masses is inevitable, the precise values of
the amplitudes depend on the assumed parameters of the accretion
column. As discussed in \S2, we chose physically motivated values
based on the assumption of spin equilibrium for this study. The
excellent agreement between the observed distribution of ms~pulsar
periods and those expected from the calculations support these
assumptions. However, somewhat different values of the column opening
angles as well as the details of the radiation beam can affect the
quantitative results. In addition, if high mass accretion rates indeed
lead to magnetic field decay, this can further attenuate pulsation
amplitudes (Cumming et al.\ 2001).

The intermittent appearance of detectable pulsations recently seen in
Aql~X-1 (Casella et al.\ 2007), HETE J1900.1-2455 (Galloway et al.\
2007), and SAX~J1748.9$-$2021 (Altamirano et al.\ 2008) may also be
related to the changing properties of the accretion disk and column
over short timescales. In particular, high mass accretion rates lead
to a larger opening angle of the accretion column, which leads to
lower pulsation amplitudes. This is true even for small fluctuations
in the mass accretion rate in the annulus where magnetic field lines
are loaded, which is shown to be variable and turbulent (Romanova,
Kulkarni, \& Lovelace 2008) and can move the loading radius around its
equilibrium position.  Alternatively, this can be a result of more
dramatic events in the accretion process, such as a thermonuclear
(Type-I) burst clearing the inner disk and moving the loading radius
outwards. If so, there may be correlated changes between the
appearance of pulsations and flux levels, spectral properties, and/or
bursting activity. The suggested correlation between the Type-I
thermonuclear bursts and the subsequent appearance of pulsations
(Galloway et al.\ 2007; Altamirano et al.\ 2008) is very intriguing in
this respect.

Finally, we address the presence of detectable flux oscillations
during thermonuclear bursts in a number of LMXBs. These so-called
burst oscillations appear during either the rise or the cooling phase
of the bursts and have rms amplitudes ranging from $\sim 1-10\%$.
Even for LMXBs that are massive enough to suppress persistent
pulsations, burst oscillations may occur with these relatively high
amplitudes because they are thought to arise from {\it one} spot as
opposed to the two caps present during persistent emission.
Specifically, the one hotspot is thought to correspond to a burning
front a mode in the burst rise, or a $m=1$ mode in the burst
tail. Pulsation amplitudes are significantly lower and fall off faster
with two caps (see, e.g., Muno et al.\ 2003), making persistent
pulsations much harder to detect than burst oscillations.

\acknowledgements

I thank Craig Markwardt for sharing unpublished data on XTE J1814-338,
Paul Ray for his help with the ASM data, Dimitrios Psaltis and Deepto
Chakrabarty for numerous useful discussions and comments on the
manuscript. I thank an anonymous referee for suggesting to look at the
correlation between spin frequencies and mass accretion rates, as well as
for other insightful comments. I also thank K. Finlator and T. G\"uver
for useful suggestions. I acknowledge support from NSF grant
AST-0708640.

\begin{deluxetable}{lcccc}
\tablecolumns{5}
\tablewidth{420pt}
\tablecaption{Average Accretion Rates of Neutron-Star Binaries}
\footnotesize
\tablehead{
Source	&	Distance &	ASM countrate &	Luminosity & $\langle
\dot{M}\rangle$\\
	&	  (kpc)  &	   (ct~$s^{-1}$) &    (erg~s$^{-1}$) &
	($M_\odot$~yr$^{-1}$)}
\startdata
GX~17+2	&		9.8 &	44.9 &	1.38$\times 10^{38}$ &	1.16$\times
10^{-8}$\\
Cyg~X-2	&		10  &	37.6 &	1.20$\times 10^{38}$ &	1.01$\times
10^{-8}$\\
Ser~X-1	&		7.7 &	16.52 &	3.13$\times 10^{37}$ &	2.63$\times
10^{-9}$\\
4U~1735--44 &		6.5 &	13.95 &	1.88$\times 10^{37}$ &	1.59$\times
10^{-9}$\\
GX~3+1	&		5 &	21.49 &	1.72$\times 10^{37}$ &	1.44$\times
10^{-9}$\\
4U~1746--37 &		16 &	2.068 &	1.69$\times 10^{37}$ &	1.42$\times
10^{-9}$\\
3A1820--303 &		4.9 &	20.7 &	1.59$\times 10^{37}$ &	1.34$\times
10^{-9}$\\
4U~1705--44 &		5.8 &	11.08 &	1.19$\times 10^{37}$ &	1.00$\times
10^{-9}$\\
4U~1636--536 &		5.9 &	9.78 &	1.11$\times 10^{37}$ &	9.31$\times
10^{-10}$\\
KS~1731--260 &		5.6 &	3.66 &	3.67$\times 10^{36}$ &	3.09$\times
10^{-10}$\\
4U~1728--34 &		4 &	6.84 &	3.49$\times 10^{36}$ &	2.94$\times
10^{-10}$\\
GRS~1741.9--2853 &	6 &	1.376 &	1.58$\times 10^{36}$ &	1.33$\times
10^{-10}$\\
MXB~1659--298 &		9 &	0.527 &	1.37$\times 10^{36}$ & 	1.15$\times
10^{-10}$\\
4U~1702--429 &		4.2 &	2.378 &	1.33$\times 10^{36}$ &	1.12$\times
10^{-10}$\\
4U~1608--52 &		3.2 &	3.76 &	1.23$\times 10^{36}$ &	1.04$\times
10^{-10}$\\
Aql~X-1	&		3.9 &	1.864 &	9.07$\times 10^{35}$ &	7.63$\times
10^{-11}$\\
4U~1916--053 &		6.8 &	0.581 &	8.59$\times 10^{35}$ &	7.23$\times
10^{-11}$\\
GRS~1747--312 &		 9 &	0.298 &	7.72$\times 10^{35}$ &	6.49$\times
10^{-11}$\\
XTEJ1710--281 &	        12 &	0.1104 &5.08$\times 10^{35}$ &	4.28$\times
10^{-11}$\\
EXO~0748--676 &		5.7 &	0.38 &	3.95$\times 10^{35}$ &	3.32$\times
10^{-11}$\\
XB~1832--330 &		6.7 &	0.1836&	2.64$\times 10^{35}$ &	2.22$\times
10^{-11}$\\
EXO~1754--248 &		4.7 &	0.363 &	2.59$\times 10^{35}$ &	2.18$\times
10^{-11}$\\
4U~1724--307 &		5 &	0.269 &	2.15$\times 10^{35}$ &	1.81$\times
10^{-11}$\\
SAX~J1747.0--2853 &	5.2 &	0.174 &	1.50$\times 10^{35}$ &	1.27$\times
10^{-11}$\\
4U~0919--54 &		4 &	0.255 &	1.30$\times 10^{35}$ &	1.10$\times
10^{-11}$\\
HETE~J1900.1--2455 & 	3.6 &	0.2842&	1.18$\times 10^{35}$ &	9.91$\times
10^{-12}$\\
SAX~J1750.8--2900 &	5.2 &	0.1169&	1.01$\times 10^{35}$ &	8.53$\times
10^{-12}$\\
XTE~J1814--338 &	7.9 &	0.0446&	8.90$\times 10^{34}$ &	7.49$\times
10^{-12}$\\
SAX~J1808.4--3658 &	2.8 &	0.118&	2.89$\times 10^{34}$ &	2.44$\times
10^{-12}$\\
4U~2129+12 &		5.8 &	0.0125&	1.34$\times 10^{34}$ &	1.13$\times
10^{-12}$\\
IGR~J00291+5934 &	4 &	0.02102&1.07$\times 10^{34}$ &	9.05$\times
10^{-13}$\\
\hline
\enddata
\end{deluxetable}

\begin{deluxetable}{lcccc}
\tablecolumns{5}
\tablewidth{420pt}
\tablecaption{Millisecond Pulsar Properties}
\tablehead{
Source	    &        Spin Freq. (Hz) & P$_{\rm orb}$ &	rms Amplitude & Reference$^a$}
\startdata
SAX J1808.4-3658   &  401   &   2.01~hr       &   4.1$-$7\%       &   1  \\      
XTE J1751-305      &  435   &   42.4~min      &   3.5\%           &   2  \\      
XTE J0929-314      &  185   &   43.6~min      &   3$-$7\%         &   3  \\ 
XTE J1807-294      &  191   &   41.0~min      &   3.1$-$5.99\%    &   4  \\
XTE J1814-338      &  314   &   4.27~hr       &   12\%            &   5  \\ 
IGR J00291+5934    &  599   &   2.46~hr       &   8\%             &   6  \\
HETE J1900.1-2455  &  377   &   83.25~min     &   1$-$2/3\%       &   7  \\
SWIFT J1756.9-2508 &  182   &   54.7~min      &   4.2 \%          &   8  \\
\enddata
\tablenotetext{a}{References.- 1. Wijnands \& van der Klis 1998;
2. Markwardt et al.\ 2002; 3.  Galloway et al.\ 2002; 4.  Kirsch et
al. 2004; 5. Watts, Strohmayer, \& Markwardt 2005; 6.  Galloway et
al.\ 2005; 7.  Galloway et al.\ 2007; 8.  Krimm et al.\ 2007.}
\end{deluxetable}


\begin{figure}
\centerline{\epsfig{file=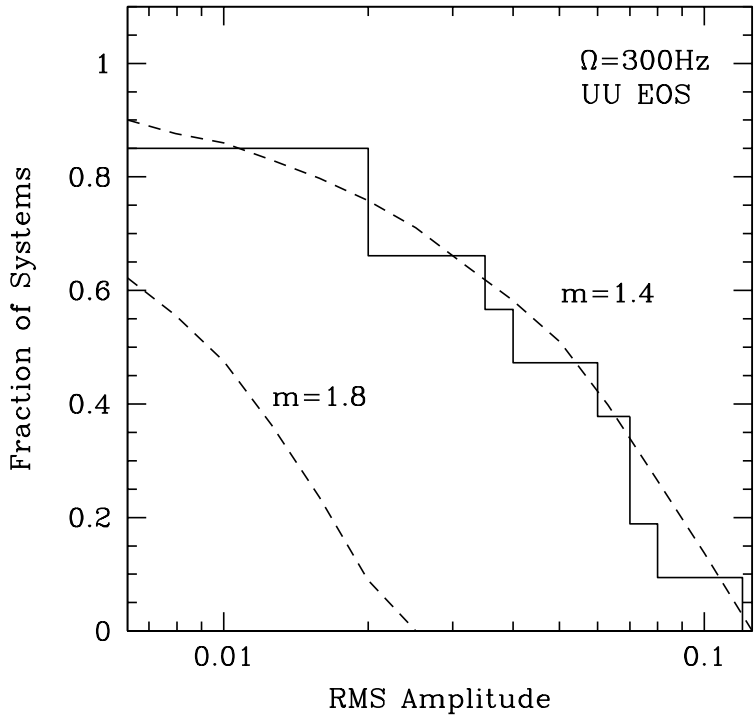,width=0.75\linewidth}}
\caption {The dashed lines show the probability that an accreting
neutron star with the specified mass ($m \equiv M/M_\odot = 1.4$ or
$M/M_\odot = 1.8$) will appear as a millisecond pulsar with a rms
amplitude given on the x-axis. The fraction of sources that appear 
as pulsars with a detectable rms amplitudes are drastically different 
between the $1.4 M_\odot$ and $1.8 M_\odot$ mass neutron stars. 
The histogram shows the rms pulse amplitudes of the eight ms~pulsars 
observed to date. The remarkable agreement between the expected 
distribution for $M=1.4 M_\odot$ neutron stars and that of the
observed distribution lend support to the idea that these neutron 
stars have not accreted a significant amount from their binary 
companion over their lifetime and that their low masses are likely 
to be responsible for observable pulsations in these systems. In 
constrast, only a very small fraction high-$<\dot{M}>$ LMXBs 
(two sources) have shown detectable pulsations at $\sim 1\%$ 
amplitude to date, as expected from the distribution for 
$M=1.8 M_\odot$ neutron stars.  } 
\label{Fig:prob} \end{figure}

\begin{figure} 
\centerline{\epsfig{file=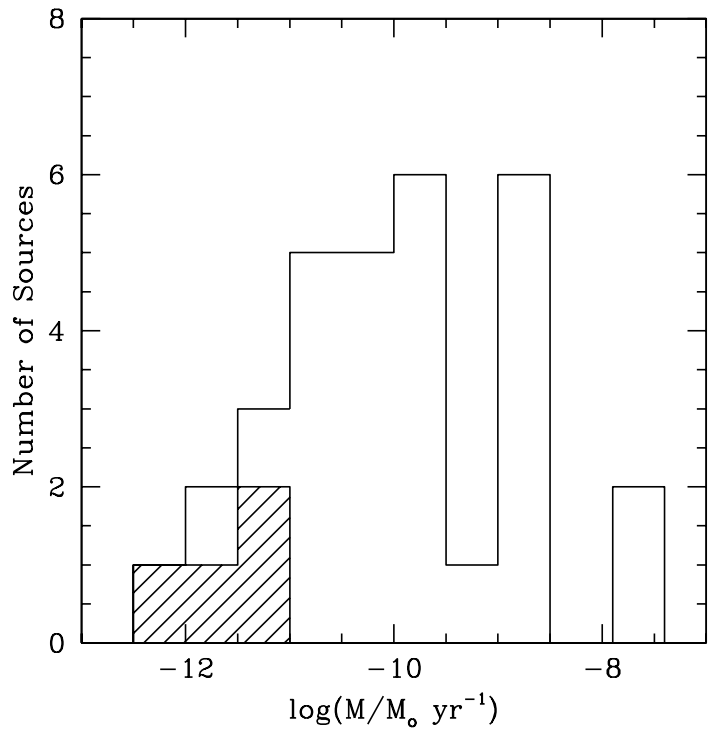,width=0.75\linewidth}}
\caption {The mass accretion rate for low-mass X-ray binaries and
ms~pulsars (hatched region) determined from the RXTE All Sky Monitor
lightcurves.  The LMXB sample is selected by the presence of
thermonuclear bursts so that a distance estimate based on Eddington
limited bursts can be made.}  
\label{Fig:mdot} 
\end{figure}

\begin{figure} 
\centerline{\epsfig{file=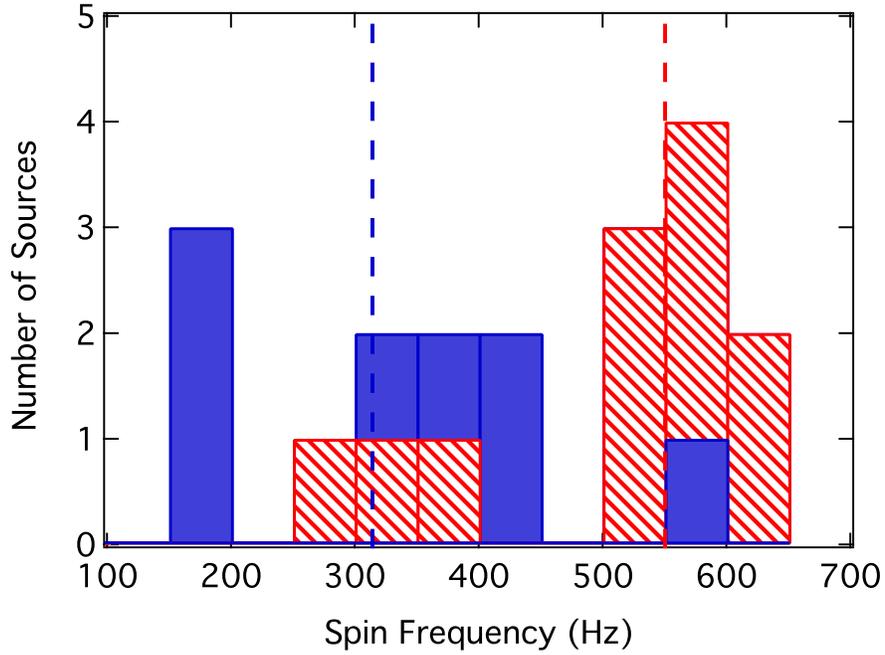,width=0.75\linewidth}}
\caption {The spin frequency distributions of the MSPs as determined
by their persistent pulsations and that of the LMXBs measured from
burst oscillations (Galloway et al.\ 2008). The dashed lines show the 
medians of the two distributions. As expected from spin equilibrium 
arguments, the low mass accretion rate systems, i.e., the MSPs, 
have a distribution of spin frequencies that is displaced 
towards lower values.}
\label{Fig:spindist} 
\end{figure}

\end{document}